\documentclass[aps,twocolumn,prl,amsfonts,amssymb,amsmath]{revtex4-1}

\usepackage[latin1,utf8]{inputenc}

\usepackage{amssymb,amsmath, amsbsy}
\usepackage{bm}% bold math
\usepackage{physics}

\usepackage{multirow}
\usepackage{graphicx}% Include figure files
\usepackage{subfigure}
\usepackage{tikz}

\usepackage{float}
\usepackage{comment}
\usepackage{blindtext}
\usepackage{tabularx}
\newcommand{\el}[1]{{\color{black} #1}}

\begin{document}
    \title{Nonequilibrium fluctuation relations for non-Gaussian processes} 
    \author{Arthur M. Faria$^{1,2}$}
    \author{Marcus V. S. Bonança$^1$}
    \author{Eric Lutz$^2$}
    \affiliation{$^1$Instituto de Física Gleb Wataghin, Universidade Estadual de Campinas,   13083-859 Campinas, São Paulo, Brazil
    \\
    $^2$
    Institute for Theoretical Physics I, University of Stuttgart, D-70550 Stuttgart, Germany%\\This line break forced with \textbackslash\textbackslash
    }%

%%%%%%%%%%%%%%%%%%%%%%%%
\begin{abstract}
Non-Gaussian noise is omnipresent in  systems where the central-limit theorem is inapplicable. We here investigate the stochastic thermodynamics of small systems that are described by a general Kramers-Moyal equation that includes both Gaussian and non-Gaussian white noise contributions. We obtain detailed and integral fluctuation relations for the nonequilibrium entropy production of  these Markov processes in the regime of weak noise. As an application, we analyze the properties of driven objects that are locally coupled to a heat bath via a finite-range interaction, by considering  an overdamped  particle that is pulled by a moving harmonic potential. We find that reducing the bath interaction range increases non-Gaussian features, and strongly suppresses the  average nonequilibrium entropy production. We further discuss a generalized detailed-balance condition.
\end{abstract}

%%%%%%%%%%%%%%%%%%%%%%%%
\maketitle

 Macroscopic thermodynamics offers a powerful framework to study equilibrium and nonequilibrium features of large  systems \cite{kon15}. In the past two decades, this formalism has been successfully extended to describe the stochastic thermodynamics of small systems, such as biomolecules, colloidal particles  and molecular motors \cite{sek10,jar11,sei12,cil13}. These microscopic objects are subjected to thermal fluctuations that can usually be neglected at the macroscale. As a result, their properties can no longer be characterized by averaged quantities alone. In order to properly account for  fluctuations, the laws of thermodynamics  have  been formulated along individual random trajectories. In particular,  nonequilibrium generalizations of  the second law in the form of fluctuation relations, that quantify the occurrence of negative entropy production events,   have been obtained \cite{sek10,jar11,sei12,cil13}. Their validity beyond linear response makes them extremely useful in the investigation of nonequilibrium phenomena. Stochastic thermodynamics has been extensively explored experimentally, both in physical and biological systems \cite{sek10,jar11,sei12,cil13}.
    
 A common assumption in stochastic thermodynamics is that the system of interest is  globally coupled to a  heat reservoir via an infinite-range interaction. According to the central-limit theorem, the thermal noise experienced by the system is then Gaussian \cite{gar97,kam07,rei16,risk89}. There is, however, a large class of microscopic objects where the bath coupling is  local, with a finite interaction range given by a typical length scale of the surroundings. Concrete examples include Brownian motion in  a scattering environment made of a gas of molecules (when the coupling is limited by  the  wavelength of the environment) \cite{gal90,dio95,gal96,vac00,horn06,vac09} and diffusion in a disordered medium (when the coupling is limited by the  spatial correlation length of the disorder potential) \cite{coh97,coh97a,ang97,bul98,coh98,gui03}. The restricted range of the bath interaction will in general lead to non-Gaussian noise \cite{far23}, and strongly affect the stochastic thermodynamics of the system. Non-Gaussian processes are widespread in science and technology   whenever the central-limit theorem does not apply \cite{gri95,weg89,gov04,ben11,bla00,pek04,met20}. However, few studies have been devoted to their stochastic thermodynamics so far: the nonequilibrium fluctuations of heat transport between a thermal bath and an athermal bath have been investigated in Refs.~\cite{kan13,kan15,kan15a,kan17}, whereas the unusual fluctuation properties of microscopic systems subjected to L\'evy and Poisson noises have been examined in Refs.~\cite{tou07,tou09a,bau09,bud12}. 
     
We here study the nonequilibrium stochastic thermodynamics of general Markov processes, described by a Kramers-Moyal equation that includes both Gaussian and non-Gaussian white noise contributions \cite{gar97,kam07,rei16,risk89}. We employ path integral techniques  \cite{wie86,cha01} to derive detailed as well as  integral fluctuation relations for the nonequilibrium entropy production for weak noise. We further illustrate our results by concretely analyzing a harmonically trapped particle, dragged at constant velocity, while being locally coupled to a heat bath. Our main observation is that a decrease of the bath interaction range strongly reduces the mean entropy production. The common assumption of global reservoir coupling therefore overestimates the average  entropy production, when the coupling is restricted by a typical length scale of the environment. Finally, we obtain a generalized detailed-balance condition \cite{gar97,kam07,rei16,risk89} for the non-Gaussian environment.

%%%%%%%%%%%%%%%%%%%%
\paragraph{Fluctuation relations for  general Markov processes.} Let us  consider a  stochastic process without memory described by a   time-homogeneous Markov function. Its sample paths $x_t$ are generically composed of a deterministic drift plus continuous (Gaussian) diffusion and discrete (Poisson-like) jump contributions \cite{gar97,kam07}. For small jumps, the probability density function  $p(x,t)$ obeys the Kramers-Moyal equation  \cite{gar97,kam07},
\begin{equation}
        \partial_t p(x,t) =  \sum_{n=1}^{\infty} \frac{(-\partial_x)^{n}}{n!}\left[ a_n(x)p(x,t)\right],
        \label{KM}
    \end{equation}
with coefficients $ a_n(x)$. Equation \eqref{KM}  involves an infinite number of higher derivatives. The usual Fokker-Planck equation \cite{gar97,kam07,rei16,risk89}, which is generally  studied in stochastic thermodynamics \cite{sek10,jar11,sei12,cil13},  is obtained by keeping the first two terms of the expansion. For a particle confined in a potential $V(x)$, globally coupled to a bath at temperature $T$ with overdamped  friction $\gamma$, the  drift and diffusion coefficients are respectively  given by $a_1(x) = -V'(x)/\gamma$ and $a_2(x) = 2k T/\gamma$, where $k$ is the Boltzmann constant. This situation corresponds to Gaussian white noise \cite{gar97,kam07}; (Poisson-like) jump contributions are here absent. Keeping higher-order terms in the Kramers-Moyal equation \eqref{KM} leads to discrete jumps and to non-Gaussian white noise. It is useful to remember that, according to Pawula's theorem, the series may be stopped at the first or second order, or must contain an infinite number of terms,   to guarantee a positive probability distribution \cite{gar97,kam07}. As commonly done \cite{kub73,mat84,kne84,tou09}, we  consider the noise to be weak, compared to the drift,  to ensure that the problem is   mathematically well-behaved. To that end, we introduce a small parameter $\epsilon$, and rescale the Kramers-Moyal coefficients according to $a_n(x) \rightarrow \epsilon^{n} a_n(x)$.

In order to study the nonequilibrium stochastic thermodynamics of an object described by Eq.~\eqref{KM}, we  drive the potential $V(x,\alpha_t)$ with an arbitrary time-dependent parameter $\alpha_t$. For a constant value, $\alpha_t=\alpha$, we assume that the probability distribution relaxes to a unique steady state, $p_{s}(x)= N \exp[-\phi(x,\alpha)]$,   where $N$ is a normalization factor.     In the limit $\epsilon \rightarrow 0$, we explicitly find (Supplemental Material \cite{sup})
        \begin{align}
        \!\phi(x,\alpha) = -\sum_{n=2}^{\infty}\frac{2^{n}}{ \epsilon n!}\int_0^x \!\!dy \frac{ a_n(y)}{ a_2^n(y)}  [V'(y,\alpha)/\gamma]^{n-1}.       
    \end{align}
    It is advantageous to distinguish   Gaussian ($n=2$) and non-Gaussian ($n\geq3$) contributions to $p_{s}(x)$, and write $\phi(x,\alpha)= \phi^{G}(x,\alpha)+\phi^{NG}(x,\alpha)$. 
The Gaussian part $\phi^{G}(x,\alpha)$ only depends on the diffusion coefficient $a_2(x)$, whereas the non-Gaussian element $\phi^{NG}(x,\alpha)$ contains all the higher-order coefficients $a_n(x)$ ($n\geq 3$).
    
%    %
    \begin{figure}[t]
    \centering
      \hspace*{.4cm} \includegraphics[width=0.45\textwidth]{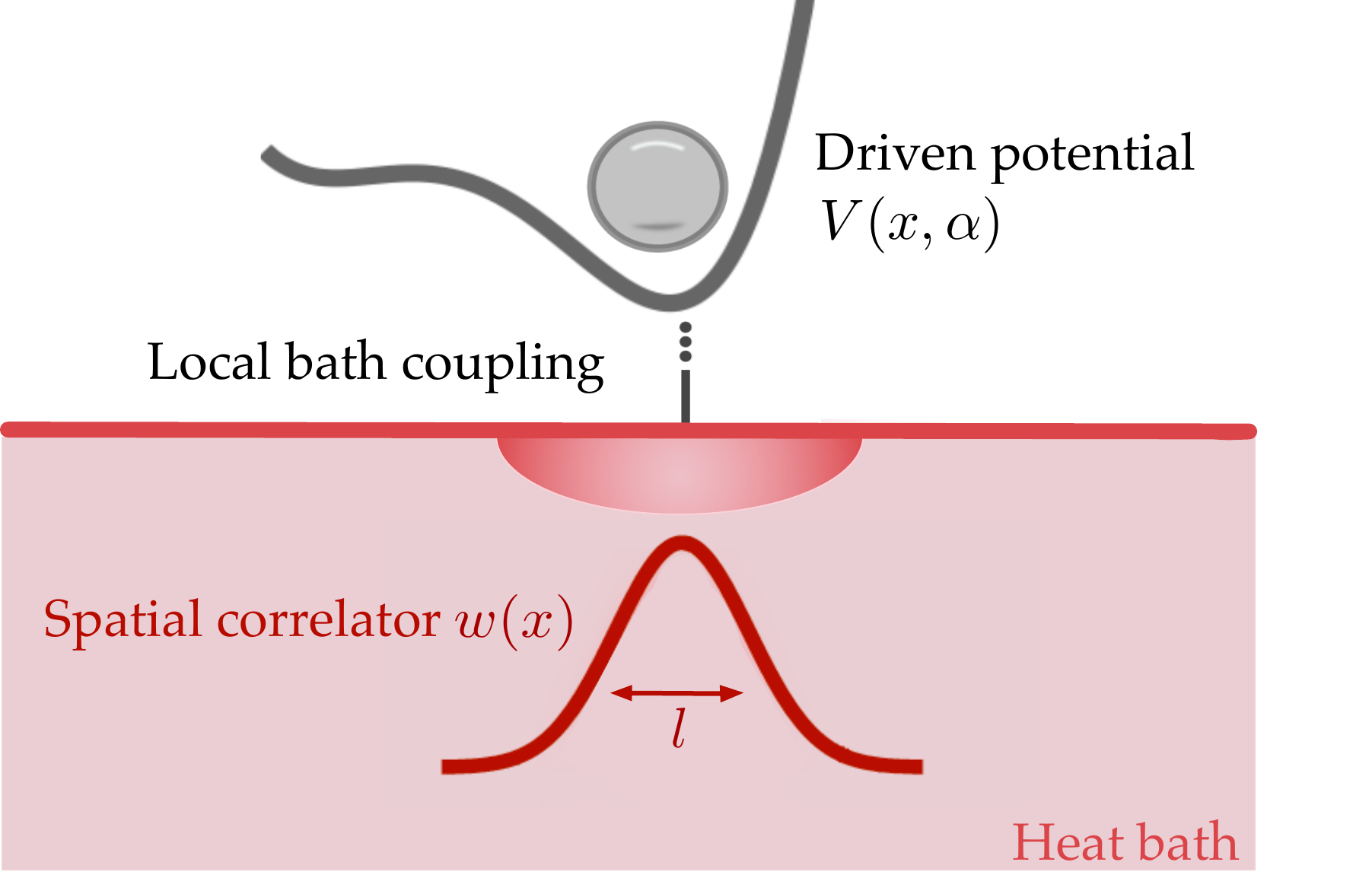}
        \caption{Schematic representation of a microscopic system  in a driven potential $V(x,\alpha$), with time-dependent driving parameter $\alpha$, locally coupled to a heat reservoir. The finite-range bath interaction is characterized by a spatial correlation function $w(x)$, with typical  length scale $l$. Global bath coupling with Gaussian noise is recovered for infinite $l$. For finite $l$, the overdamped dynamics exhibits non-Gaussian fluctuations.}
        \label{fig1}
    \end{figure}
%    %
 \begin{figure*}[t]
	\centering
	\begin{tikzpicture}
            \node (a) [label={[label distance=-.65 cm]145:\textbf{a)}}]  at (-7.2,0) {\includegraphics[width=0.31\textwidth]{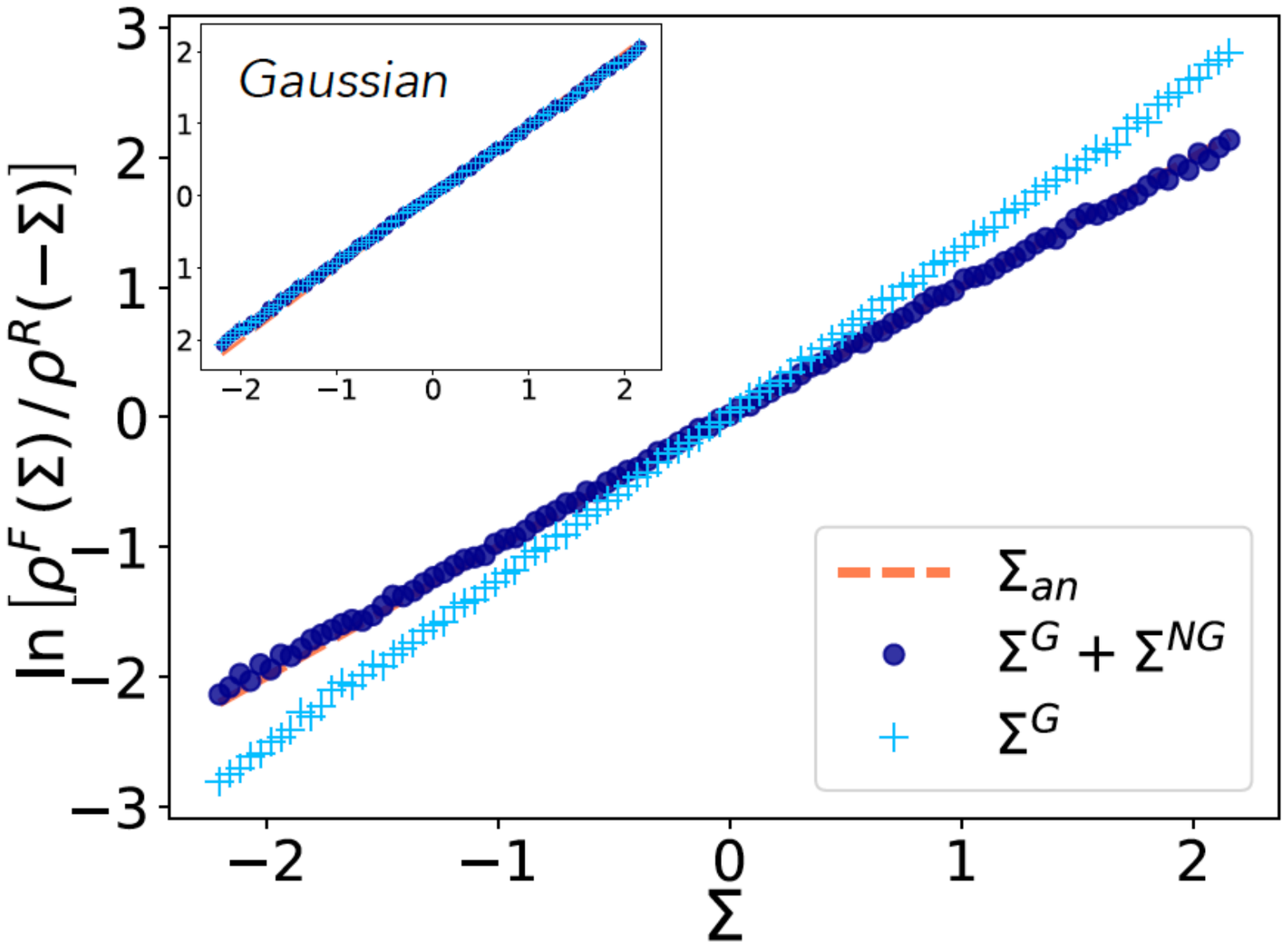}};
            \node (a) [label={[label distance=-.25 cm]148:\textbf{b)}}]  at (-1.0,0) {\includegraphics[width=0.31\textwidth]{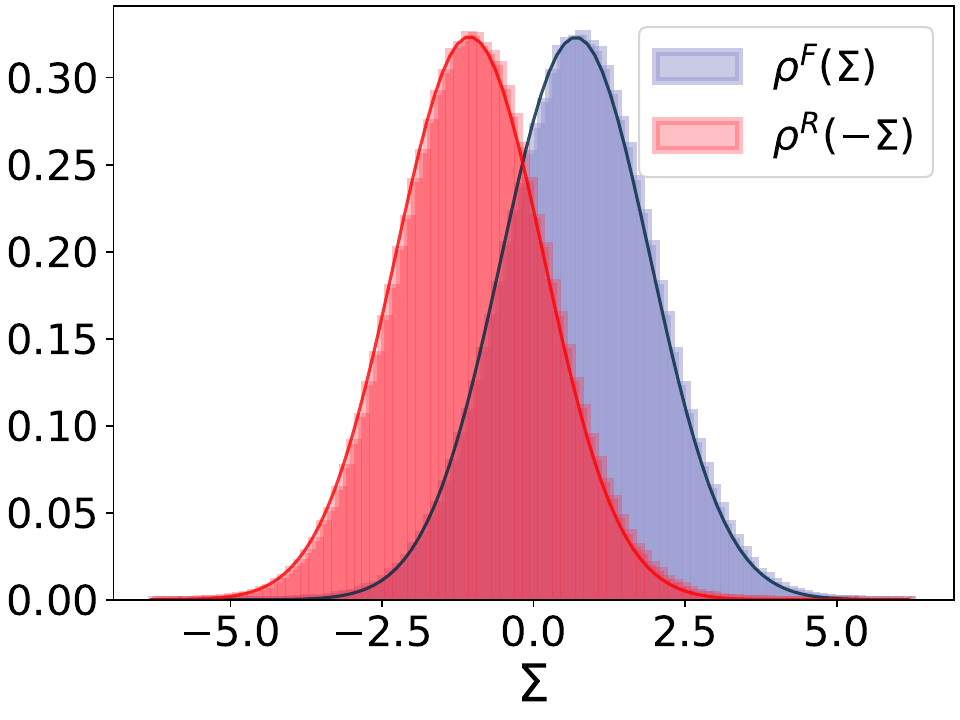}};
            \node (a) [label={[label distance=-.65 cm]145:\textbf{c)}}]  at (4.8,0) {\includegraphics[width=0.31\textwidth]{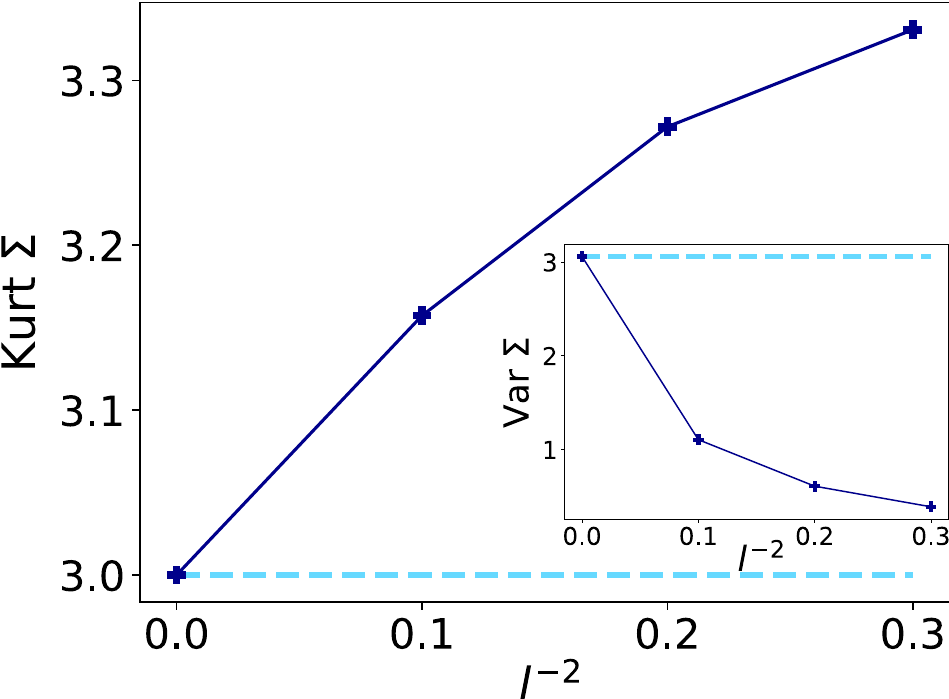}};
	\end{tikzpicture}
        \caption{Stochastic thermodynamics of a dragged harmonic particle with local bath coupling. a) The detailed fluctuation relation $\rho^{F}(\Sigma) /\rho^{R}(-\Sigma)                = e^\Sigma $, Eq.~\eqref{10}, is verified for the total entropy production, $\Sigma = \Sigma^G + \Sigma^{NG}$, that includes both Gaussian and non-Gaussian contributions  (dark blue dots), for a finite-range bath coupling length $l^{-2} = 0.05$. It is violated when only the Gaussian term $ \Sigma^G$ is considered (light blue crosses). The simulated  entropy production $\Sigma$ perfectly agrees  with the analytical expression \eqref{eq:EP_numerics} (dashed red line). The inset shows the case of a global bath interaction ($l^{-2} = 0$) that corresponds to pure Gaussian noise. b) Forward and reversed  probability distributions, $\rho^{F}(\Sigma)$ and     $\rho^{R}(-\Sigma)$ (red and blue histograms), as well as their non-Gaussian fits (red and blue solid  lines). c) The kurtosis $\text{Kurt}\,\Sigma$ (dark blue crosses) increases when the length  of the bath coupling is reduced, indicating that non-Gaussian features are accentuated, while  the variance $\text{Var}\,\Sigma$  (dark blue crosses, inset) decreases, compared to their respective Gaussian values (light blue dashed lines). Parameters are $kT = 0.5$, $\kappa = 1.0$, $\gamma=10$ and $v_0= 0.03$. The total simulation time is $\tau= 2000/\gamma$ with increment $dt=0.01$.}
        \label{fig:FT}
    \end{figure*}
    We   next derive detailed and integral fluctuation theorems for the nonequilibrium entropy production by using a path integral representation of the solution of Eq.~\eqref{KM} \cite{wie86,cha01}. We denote by $X = \{x\}^{\tau}_{-\tau}$ a trajectory of the driven system that starts at $t = -\tau$ and ends at $t = +\tau$. We furthermore consider a forward process $\alpha_t^F$ for which the parameter $\alpha_t$ is varied from an initial value $\alpha_{-\tau}^F=\alpha_i$ to a final value $\alpha_{+\tau}^F = \alpha_f$. By extending the analysis of Refs.~\cite{che06,def11} from a Fokker-Planck dynamics to the full Kramers-Moyal expansion \eqref{KM}, the conditional probability $P^F[X|x_{-\tau}]$ of observing a trajectory $X$ given the initial microstate $x_{-\tau}$ may  be expressed as 
    \begin{align}
        \nonumber P^F[X|x_{-\tau}] =    \mathcal{N} \exp\left[-\int_{-\tau}^{\tau}dt'\,S(x_t, \dot{x}_t;  \alpha_t^F)\right],
           \end{align}
   with a constant $\mathcal{N}$ and the   Onsager-Machlup  functional $S$ \cite{ons93}.  We concretely obtain    (Supplemental Material \cite{sup}) %
    \begin{align}
   \!  \!  \!S(x_t, \dot{x}_t;  \alpha_t^F) = \!\sum_{n=3}^{\infty} \frac{ a_{n}(x_t)}{ n!} \left[\frac{\dot{x}_{t} + V'(x_t,\alpha_t^F)/\gamma}{\epsilon a_2(x_t)}\right]^n\hspace{-0.5mm}.\hspace{-2mm}
    \end{align}
    It is again convenient  to write  $S$ as a sum of  a Gaussian ($n=2$) and a non-Gaussian ($n\geq3$) term,   $S = S^G + S^{NG}$.
 We may likewise consider the time-reversed process with driving function $\alpha_t^R = -\alpha_{-t}^F$. The associated  conditional probability, $P^R[X^{\dagger}|x^{\dagger}_{-\tau}]$, for the time-reversed trajectory $X^{\dagger} = \{x^{\dagger} \}^{\tau}_{-\tau}$, with the  variable $x^{\dagger}_t= x_{-t} $, reads    %
    \begin{align}
        P^R[X^{\dagger}|x^{\dagger}_{-\tau}]= & \mathcal{N} \exp\left[-\int_{-\tau}^{\tau}dt'\,S^{\dagger}(x_t, \dot{x}_t;  \alpha_t^F)\right],
    \end{align}
    where we have defined the conjugate Onsager-Machlup functional $S^{\dagger}(x_t, \dot{x}_t ;  \alpha_t) = S(x_t, -\dot{x}_t;  \alpha_t)$. The ratio of the two conditional probabilities is accordingly given by    %
    \begin{align}
    \label{eq:r_prop}
        \frac{P^F[X|x_{-\tau}]}{P^R[X^{\dagger}|x^{\dagger}_{-\tau}]} &\underset{\epsilon \to 0}{=}  \exp\left[-\frac{2}{\epsilon}\int_{-\tau}^{\tau}dt\, \dot{x}_{t} \frac{V'/\gamma}{  a_2} \right]
        \\
          &\hspace{-6mm}\times {\exp\left[-\frac{2}{\epsilon}\sum_{n=3}^{\infty} \frac{n}{ n!}\int_{-\tau}^{\tau}dt\, \dot{x}_t \frac{ a_{n}}{ a_2^{n}}   (V'/\gamma)^{n-1}\right]}\hspace{-0.5mm}, \hspace{-1mm} \nonumber 
    \end{align}
  where higher orders in $\epsilon$ have been neglected (Supplemental Material \cite{sup}).  Assuming that the system is initially in the steady state $p_{s}(x_{-\tau}, \alpha_i)$, the net probability of observing the trajectory $X$ for the forward process is   $P^{F}(X)= p_{s}(x_{-\tau}, \alpha_i)P^{F}[X|x_{-\tau}]$. For the reversed process, we similarly have  $P^{R}(X^{\dagger})= p_{s}(x_{\tau},  \alpha_f)P^{R}[X^{\dagger}|x^{\dagger}_{-\tau}]$. 
  Further noting that $p_{s}(x_{-\tau}, \alpha_i)/p_{s}(x_{\tau},  \alpha_f)= \exp\left(\Delta \phi \right)$, with $\Delta \phi = \int_{-\tau}^{\tau} dt\, \dot{\alpha}_t^F \partial_{\alpha}\phi + \dot{x}_t \partial_x \phi$, we arrive at  the ratio   %
    \begin{align}
        \frac{P^{F}(X)}{P^{R}(X^{\dagger})} &= \frac{p_{s}(x_{-\tau}, \alpha_i)P^{F}[X|x_{-\tau}]}{p_{s}(x_{\tau},  \alpha_f)P^{R}[X^{\dagger}|x^{\dagger}_{-\tau}]} = e^\Sigma,
        \label{eq:pre_SFT}
    \end{align} 
with the random nonequilibrium entropy production (Supplemental Material \cite{sup}) %
    \begin{align}
        \Sigma =  -\int_{-\tau}^{\tau} dt\, \dot{\alpha}^F_t \frac{\partial_{\alpha}p_{s}}{p_{s}}  + \int_{-\tau}^{\tau} dt\, \dot{x}_t \partial_x \phi^{NG}.
        \label{eq:Y}
    \end{align}
    We may  eventually obtain a  detailed fluctuation relation by introducing the distribution $\rho^{F}(\Sigma)=\int {\cal D}X P^F(X)\delta(\Sigma-\Sigma^F(X)) $ for an ensemble of forward process realizations. Using Eq.~\eqref{eq:pre_SFT}, we find
     \begin{align}
     \label{10}
      \rho^{F}(\Sigma)                 = e^\Sigma \rho^{R}(-\Sigma) ,
    \end{align}
with $\rho^{R}(-\Sigma) = \int {\cal D}X^{\dagger} P^R(X^{\dagger})\delta(\Sigma+\Sigma^R(X^{\dagger})) $. An integral fluctuation relation follows, as usual, after integration over the forward distribution, $\langle \exp(-\Sigma)\rangle = 1$, which directly implies  $\langle \Sigma\rangle \geq 0$ via Jensen's inequality. The above expressions are generalizations of the standard nonequilibrium equalities by Crooks \cite{cro99} and by Jarzynski \cite{jar97}  for general driven Markov processes described by the full Kramers-Moyal equation \eqref{KM} in the low-noise limit; they include  contributions of both Gaussian and non-Gaussian (white noise) fluctuations.

%%%%%%%%%%%%%%%%%%%%   
   
\paragraph{Dragged harmonic potential with local bath coupling.} The fluctuation relation \eqref{10} is valid for generic potentials $V(x,\alpha_t)$, nonequilibrium driving protocols $\alpha_t$ and (small) Kramers-Moyal coefficients $a_n(x)$. As an illustration, we now consider a harmonically trapped particle, dragged with constant velocity $v_0$ through a medium, $V(x,\alpha_t) = \kappa (x- v_0t)^2/2$, with  $\alpha_t = v_0t$ and $\kappa$  the spring constant. This system has played an important role in the first experimental investigations of fluctuation theorems using colloidal particles moving in a viscous fluid \cite{wan02,tre04} (see also Refs.~\cite{zon03,zon03a,rei04,bai06,ast07}). In contrast to these studies, we consider a particle that interacts with its environment through a local interaction that is characterized by a spatial correlation function, $w(x-x')$, with a finite correlation length $l$ \cite{gal90,dio95,gal96,vac00,horn06,vac09,coh97,coh97a,ang97,bul98,coh98,gui03}. For concreteness, we assume a Gaussian distribution of the form, $w(x-x')= l^2\exp[-(x-x')^2/2l^2]$. The length scale $l$ determines the finite-range of the coupling to the bath; global coupling (and, accordingly, Gaussian thermal white noise) is recovered in the limit $l\rightarrow \infty$. The overdamped dynamics of the driven particle is described by the Kramers-Moyal equation \cite{far23}
\begin{align}
     \partial_tp(x,t) = \frac{1}{\gamma}\partial_x[V'(x,\alpha_t) p] + \sum_{n=1}^{\infty}\! \frac{a_{2n}}{(2n)!} \partial_x^{2n} p,
        \label{eq:KM_nonG}
    \end{align}
with  coefficients $a_{2n}= \gamma(1 + 2 kT l^2)(2n)!/[2^n (\gamma l)^{2n} n!]$   \cite{far23}. Note that only even terms appear. Equation \eqref{eq:KM_nonG} reduces to the standard Ornstein-Uhlenbeck process \cite{gar97,kam07,rei16,risk89} for an infinite-range bath interaction $l\rightarrow \infty$.  
For finite $l$, the stochastic  entropy production \eqref{eq:Y} explicitly reads
\begin{align}
    &\hspace{-1mm}\Sigma =  {-\int_{-\tau}^{\tau} dt\, \dot{ a}_t^F \frac{\partial_{\alpha}p_{s}}{p_{s}} + \int_{-\tau}^{\tau} dt\, \dot{x}_t \partial_x \phi^{NG}} 
    \\
    &\hspace{-1mm}{=\int_{-\tau}^{\tau} dt\left[\frac{2\kappa}{a_2}v_0 z_t -  \sum_{n=2}^{\infty} \frac{(2\kappa)^{2n}}{2n!}\frac{a_{2n}}{a_2^{2n}} \dot z_t(v_0 + z_t^{2n-1})\right]}, \hspace{-2mm}
    \label{eq:EP_numerics}
\end{align}   where we have introduced the  variable $   z_t = x_t - v_0t$. The total entropy production can thus be written as a sum, $\Sigma = \Sigma^G + \Sigma^{NG}$, of Gaussian (non-Gaussian) contributions given by the first two (higher) terms in Eq.~\eqref{eq:EP_numerics}.

\begin{figure}[t]
    \centering
       \includegraphics[width=0.44\textwidth]{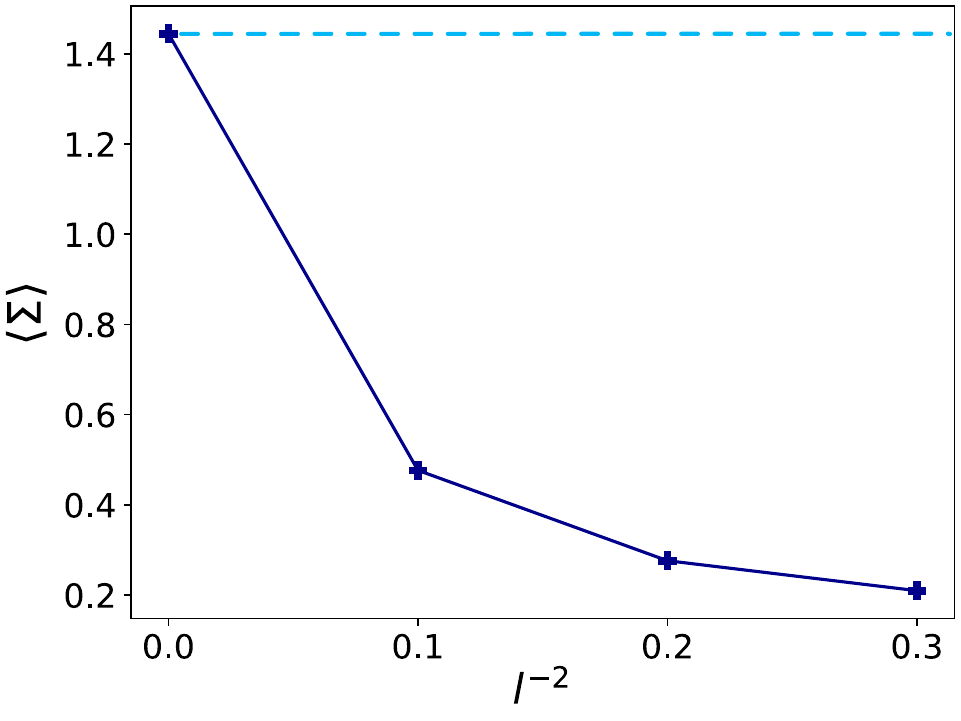}
        \caption{Average nonequilibrium entropy production. The mean entropy production $\langle \Sigma \rangle$, average of Eq.~\eqref{eq:EP_numerics} (dark blue crosses), for the dragged harmonic potential decreases when the bath coupling length scale $l$  is   reduced. The assumption of global bath coupling (light blue dashed line) thus overestimates the nonequilibrium entropy production for a finite reservoir interaction range. Same parameters as in Fig.~2.}
        \label{fig3}
    \end{figure}
    
Figure 2 displays the results of a numerical simulation of the non-Gaussian process described by Eq.~\eqref{eq:KM_nonG} for an ensemble of 10$^6$ trajectories (Supplemental Material \cite{sup}). Figure 2a)  confirms that the integral fluctuation relation \eqref{10} is satisfied for the total entropy production, $\Sigma = \Sigma^G + \Sigma^{NG}$, for a finite value of the length $l^{-2}=0.05$ (dark blue dots), in agreement with the analytical expression \eqref{eq:EP_numerics} (dashed red line). By contrast, the fluctuation theorem is violated when only the Gaussian contribution $\Sigma^G$ is considered (light blue crosses); the standard Gaussian case (global bath coupling with $l^{-2}=0$) is displayed in the inset. Figure 2b) further shows the respective forward and reversed probability distributions, $\rho^{F}(\Sigma)$ and $\rho^{R}(-\Sigma)$, (histograms) as well as their non-Gaussian fits (solid lines), for  local bath coupling ($l^{-2}=0.05$). Figure~2c) additionally represents the kurtosis of the forward distribution, $\text{Kurt} \,\Sigma = \langle  (\Sigma -\langle \Sigma\rangle)^4 \rangle/ [\langle (\Sigma -\langle \Sigma \rangle )^2\rangle ]^2$,  a measure of the deviation from a normal distribution \cite{gar97,kam07,rei16,risk89} (dark blue symbols), as well as the variance, $\text{Var}\, \Sigma = (\Sigma -\langle \Sigma\rangle)^2$, a measure of the width of the distribution (inset), as a function of the inverse length scale $l$. We observe that  the kurtosis increases whereas the variance decreases compared to their Gaussian values (light blue dashed lines), when the length $l$ decreases. This indicates that the distribution $\rho^{F}(\Sigma)$ becomes narrower and more non-Gaussian, when the reservoir coupling range is reduced. 

Let us now  analyze the thermodynamic consequences of  a finite-range bath interaction. Figure 3 indicates that  the mean nonequilibrium entropy production $\langle \Sigma \rangle$ is sharply suppressed (dark blue symbols) when the bath-coupling range is decreased compared to the situation of a global reservoir interaction.   This behavior may be physically understood by noting that the system would effectively not be coupled to any  reservoir in the extreme limit of zero-range interaction $(l \rightarrow 0)$, which would imply vanishing average entropy production \el{(and, in turn, vanishing variance, since the two quantities are almost proportional \cite{com})}. These findings convey the important message that the assumption of global bath coupling overestimates the nonequilibrium entropy production when the typical length scale of the environment limits the interaction range.
 
\begin{figure}[t]
        \centering
        \includegraphics[width=0.44\textwidth]{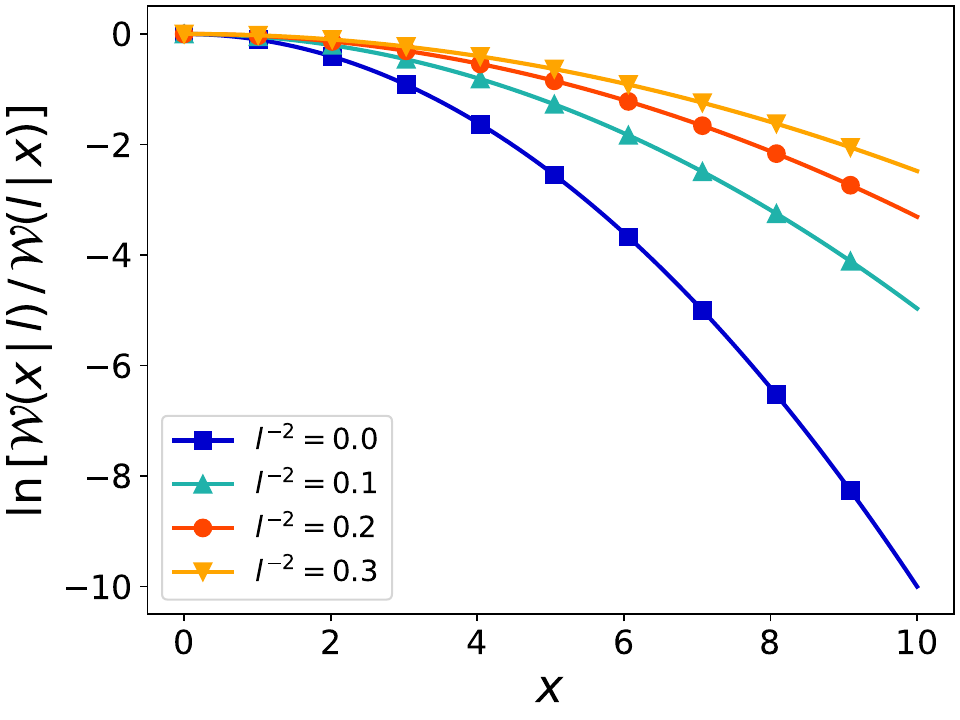}
        \caption{Detailed balance condition. Reducing the finite reservoir interaction range $l$ modifies the detailed balance condition of the dragged harmonic potential by increasing  the ratio of the transition probabilities, $\ln[{W(x|x')}/{W(x'|x)}]$, Eq.~\eqref{14},  with $x'=x+1/l^2$. Same parameters as in Fig.~2.}
    \end{figure}

%%%%%%%%%%%%%%%%%%%%
\paragraph{Modified detailed-balance condition.} Since  the probability current vanishes in a stationary state, individual transition probabilities (from and to a given state) must balance themselves. This key property of a stationary system is known as detailed-balance condition  \cite{gar97,kam07,rei16,risk89}. The presence of non-Gaussian  noise will affect that condition. To see this, we write the evolution of the density probability $p(x,t)$ of a  general Markov process  in the form of a differential Chapman-Kolmogorov equation \cite{gar97,kam07}
\begin{eqnarray}
\!\!\!\!\partial_t p(x,t) &= &\!-\partial_x [a_1(x) p(x,t)] +\frac{1}{2}\partial^2_x [a_2(x) p(x,t)] \nonumber \\
&+&\!\!\int dx' [W(x|x') p(x',t) - W(x'|x)p(x,t)],
\end{eqnarray}
 with transition probabilities $W(x|x')$ and $W(x'|x)$. Stationarity, $\partial_t p_s(x) = 0$, then implies the detailed balance equation $ W(x|x')p_s(x') = W(x'|x)p_s(x)$  for the transition probabilities \cite{gar97,kam07,rei16,risk89}. This leads to
 \begin{equation}
 \label{14}
 \frac{W(x|x')}{W(x'|x)} = e^{-[\phi^{G}(x) - \phi^{G}(x')]} e^{-[\phi^{NG}(x) - \phi^{NG}(x')]},
 \end{equation}
with the stationary probability distribution $p_s(x) = N \exp[-\phi^G(x)-\phi^{NG}(x)]$. For a general Markov process, the detailed-balance condition hence includes a contribution from the non-Gaussian potential $\phi^{NG}(x)$. We note that    $W(x|x')$ is related to the dynamic structure factor of the environment \cite{vac09}, that characterizes its density fluctuations. It is thus directly accessible in a suitable scattering experiment \cite{han13,pot10}, and would allow one to determine the characteristics of the non-Gaussian reservoir. Figure 4 shows the function $\ln[{W(x|x')}/{W(x'|x)}]$ for the example of the dragged harmonic particle \eqref{eq:KM_nonG}, with $x'=x+1/l^2$, a stochastic jump from $x$ away, as a function of $x$ for several values of the length scale $l$. We note that non-Gaussian noise increases the ratio of the transition probabilities, compared to the Gaussian case.

%%%%%%%%%%%%%%%%%%%%%%%%
\paragraph{Conclusions.}  The coupling to  a Gaussian heat bath is in many cases  an idealized approximation that  breaks down when the central-limit theorem does not hold. Using a path integral (Onsager-Machlup) method, we have derived nonequilibrium fluctuation relations for general non-Gaussian Markov processes that are described by a full Kramers-Moyal equation in the weak noise regime. We have applied these findings to examine the thermodynamic consequences of a finite-range bath interaction. We have found that a reduced reservoir coupling range enhances non-Gaussianity and, at the same time, suppresses the average entropy production below its Gaussian value. It additionally affects the detailed-balance condition.  Our results extend stochastic thermodynamics to non-Gaussian Markov processes, and signal that the assumption of global bath coupling overestimates irreversible dissipation for small bath correlation lengths.

%%%%%%%%%%%%%%%%%%%%%%%%
\vspace{2mm}
    \paragraph{Acknowledgments.} The authors  acknowledge  financial support from the Brazilian agency CNPq (Grant Nos. 142556/2018-1 and 304120/2022-7), FAPESP (Grant No. 2020/02170-4), and from the German Science Foundation DFG (Grant No. FOR 2724).

%%%%%%%%%%%%%%%%%%%%%%%%%%%%%%%%%%%%%%%%%%%%%%%%%%%%%%%%%%%%%%%%%%%%%%%%%%%%%%%%%%%%%%%%%%%%%%

%%%%%%%%%%%%%%%%%%%%%%%%%%%%%%%%%%%%%%%%%%%%%%%%%%%%%%%%%%%%%%%%%%%%%%%%%%%%%%%%%%%%%%%%%%%%%%
\pagebreak
\widetext
%%%%%%%%%% Merge with supplemental materials %%%%%%%%%%
\setcounter{equation}{0}
\setcounter{figure}{0}
\setcounter{table}{0}
\setcounter{page}{1}
\makeatletter
\renewcommand{\theequation}{S\arabic{equation}}
\renewcommand{\thefigure}{S\arabic{figure}}
\renewcommand{\bibnumfmt}[1]{[S#1]}
\renewcommand{\citenumfont}[1]{S#1}

\newpage 
\begin{center}
\vskip0.5cm
{\large \bf Supplemental Material: Nonequilibrium fluctuation relations for non-Gaussian processes}
\end{center}
 \title{Nonequilibrium fluctuation relations for non-Gaussian processes} 
  
  The Supplemental Material contains details of the derivation of (I) the stationary state of the Kramers-Moyal equation in the weak-noise regime and of (II) the detailed fluctuation relation, as well as (III) a discussion of the numerical simulation of the Kramers-Moyal equation.
  %%%%%%%%%%%%%%%%%%%%
\section{I.  Steady State of the Kramers-Moyal equation in the low-noise regime}
In this Section, we compute the steady-state solution of the Kramers-Moyal equation in the small-noise  regime. To that end, we write the equation in the form $\partial_t p(x,t) = \mathcal{L}\left[p(x,t)\right]$, where we have introduced the Liouville operator $\mathcal{L}$ \cite{gar97s,kam07s,rei16s,risk89s}. The steady-state solution $ p_{s}$ then satisfies
    \begin{align}
        \mathcal{L}\left[p_{s}\right] &= - \epsilon a_1(x)p_{s} + \sum_{n=2}^{\infty}\frac{(-\epsilon)^n}{n!} \partial_x^{n-1}\left[  a_n(x) p_{s}\right]= 0
        \label{eq:ss_master}
    \end{align}
 In the low noise limit, $\epsilon \to 0$,   we consider the  ansatz \cite{kub73s,mat84s,kne84s,tou09s,kan13s}     
    \begin{align}
         p_{s} = e^{\phi_0 + \phi_1} \underset{\epsilon \to 0}{\sim} e^{\phi_0} \left(1 +\phi_1\right) = P_0 + P_1,
        \label{eq:ansatz}
    \end{align}
    \noindent where $P_0 = e^{\phi_0} $ and $P_1 = e^{\phi_0}\phi_1 $ are the respective nonperturbative and  perturbative contributions to the steady-state distribution. As a result, we may write $\mathcal{L}= \mathcal{L}_0 + \mathcal{L}_1 $ with
    \begin{align}
        &\mathcal{L}_0 [P_0] = - \epsilon a_1(x)P_0 + \frac{\epsilon^2}{2}\partial_x \left[ a_2(x)P_0\right] =0, 
        \\
        &\mathcal{L}_ 1 [P_1] =  \sum_{n=3}^{\infty} \frac{(-\epsilon)^n}{n!}\partial_x^{n-1}\left[ a_n(x)P_1\right] = 0,
        \\
        &\mathcal{L}  [p_{s}] = \mathcal{L}_0 P_1 + \mathcal{L}_ 1 P_0  = 0.
    \end{align}
 These equations can be recursively solved, with explicit solutions 
    \begin{align}
        &P_0 = \exp(\int_0^x dy \frac{\epsilon a_1(y) - \frac{\epsilon^2}{2}\partial_y  a_2(y)}{\frac{\epsilon^2}{2}a_2(y)} + C_1),
        \\
        &P_1 = P_0\left(C_2  - \sum_{n=3}^{\infty}\frac{(-\epsilon)^n}{n!}\int_0^x dy \frac{1}{\frac{\epsilon^2}{2} a_2(y)}  \frac{\partial_y^{n-1}\left[ a_n(y)P_0\right]}{P_0}\right),
    \end{align}
     \noindent where $C_1$ and $C_2$ are constants determined respectively by $\int_{-\infty}^{\infty} dy P_0(y)=1$ and $\int_{-\infty}^{\infty} dy P_1(y)=0$. We may  set the constant $C_2=0$ as it does not affect the derivation of the fluctuation theorem. We therefore obtain
     \begin{align}
     \label{s8}
         p_{s}(x) = N e^{\int_0^x dy \frac{2 \epsilon a_1(y) - \epsilon^2\partial_y  a_2(y)}{\epsilon^2 a_2(y)}}\left(1   - \sum_{n=3}^{\infty}\frac{(-\epsilon)^{n}}{n!}\int_0^x dy \frac{2}{\epsilon^2 a_2(y)}  \frac{\partial_y^{n-1}\left[ a_n(y)P_0\right]}{P_0}\right),
    \end{align}
     \noindent with $N = e^{C_1}$ a normalization factor. We additionally  note that in the low-noise limit, we  have
    \begin{align}
        &\partial_x^2\left[ a_3P_0\right] =  a_3 \partial_x^2\left[P_0\right] + 2\partial_x\left[ a_3\right]  \partial_x \left[P_0\right] +  P_0 \partial_x^2\left[ a_3\right] \ \underset{\epsilon \to 0}{\sim} \  a_3 \partial_x^2\left[P_0\right],
        \\
        &\partial_x^3\left[ a_4P_0\right] =  a_4 \partial_x^3\left[P_0\right] + 3\partial_x\left[ a_4\right]  \partial_x^2 \left[P_0\right] + 3\partial_x^2\left[ a_4\right]  \partial_x \left[P_0\right] +  P_0 \partial_x^3\left[ a_4\right] \ \underset{\epsilon \to 0}{\sim} \  a_4 \partial_x^3\left[P_0\right],
        \\
        \nonumber & \hspace{15mm}\vdots
        \\
        &\partial_x^{n}\left[ a_n P_0\right] =\sum_{k=0}^{n} \begin{pmatrix} n\\ k \end{pmatrix}   a_n^{(n-k)}(x)P_0^{(k)} \ \underset{\epsilon \to 0}{\sim} \  a_n \partial_x^{n}\left[P_0\right], 
    \end{align}
    since 
    \begin{align}
        &\partial_x \left[P_0\right] =  \frac{ 2\epsilon a_1(x) - \epsilon^2\partial_x  a_2(x)}{\epsilon^2  a_2(x)} P_0 \ \underset{\epsilon \to 0}{\sim}  \ \frac{2 a_1(x)}{\epsilon a_2(x)}P_0,
        \\
        & \partial_x^2 \left[P_0\right] = \left(\partial_x\left[\frac{2 \epsilon a_1(x) - \epsilon^2\partial_x  a_2(x)}{\epsilon^2 a_2(x)}\right] + \left(\frac{2 \epsilon a_1(x) - \epsilon^2\partial_x  a_2(x)}{\epsilon^2  a_2(x)}\right)^2 \right)P_0 \ \underset{\epsilon \to 0}{\sim} \ \left(\frac{2 a_1(x)}{\epsilon  a_2(x)}\right)^2 P_0, 
        \\
        \nonumber & \hspace{15mm}\vdots
        \\
        &\partial_x^n \left[P_0\right] \underset{\epsilon \to 0}{\sim} \ \left(\frac{2 a_1(x)}{\epsilon  a_2(x)}\right)^n P_0,
    \end{align}
or, in other words,  $ \ \epsilon^n\partial_x^n \left[P_0\right]\, \sim \, \mathcal{O}(\epsilon^{0})$.  Equation~\eqref{eq:ss_master} may then by expressed as   
    \begin{align}
        \mathcal{L}\left[ p_{s}\right] = 0 &= - \epsilon a_1(x) p_{s}  + \frac{\epsilon^2}{2} \partial_x\left[  a_2(x) p_{s}\right] + \sum_{n=3}^{\infty}\frac{(-\epsilon)^n}{n!}\partial_x^{n-1}\left[  a_n(x) p_{s}\right],
        \\
        &\underset{\epsilon \to 0}{\sim}  - \epsilon a_1(x) p_{s} + \frac{\epsilon^2}{2} a_2(x) \partial_x\left[  p_{s}\right] + \sum_{n=3}\frac{(-\epsilon)^n}{n!} a_n(x) \partial_x^{n-1}\left[  p_{s}\right].
    \end{align}
        Equation \eqref{s8} may accordingly  be simplified to
        \begin{align}
        \nonumber p_{s}(x) &\underset{\epsilon \to 0}{\sim}N e^{\int_0^x dy \frac{2 a_1(y) }{\epsilon a_2(y)}}\left(1  - \sum_{n=3}^{\infty}\frac{(-\epsilon)^n}{n!}\int_0^x dy \frac{2}{\epsilon^2 a_2(y)}  \frac{\partial_y^{n-1}\left[ a_n(y)P_0\right]}{P_0}\right)
        \\
        &= N \exp(\frac{2}{\epsilon}\int_0^x dy \frac{ a_1(y)}{ a_2(y)} + \frac{2}{\epsilon}\sum_{n=3}^{\infty}(-1)^{n-1}\frac{2^{n-1}}{ n!}\int_0^x dy \frac{ a_n(y)}{ a_2^n(y)}   a_1^{n-1}(y)).
        \label{eq:steady_state}
    \end{align}
Introducing now the decomposition, $\phi(x) = \phi^{G}(x) + \phi^{NG}(x)$, where $\phi^{G}(x)$ and $\phi^{NG}(x)$ represent the respective Gaussian and non-Gaussian steady-state contributions, we find
        \begin{align}
         p_{s}(x) &\underset{\epsilon \to 0}{\sim} N \exp[-\phi(x)] = {N} \exp[-\phi^{G}(x)] \exp[-\phi^{NG}(x)].
    \end{align}
   with 
       \begin{align}
        &\phi^{G}(x,\alpha) = \frac{2}{\epsilon}\int_0^x dy\frac{1}{a_2(y)}  V'(y,\alpha)/\gamma,
        \\
        &\phi^{NG}(x,\alpha) = -\frac{2}{\epsilon}\sum_{n=3}^{\infty}\frac{2^{n-1}}{ n!}\int_0^x dy \frac{ a_n(y)}{a_2^n(y)} \left(V'(y,\alpha)/\gamma\right)^{n-1},
    \end{align}
    where  we have taken  $a_1(x) = -V'(x, \alpha)/\gamma$, with the confining potential $V(x, \alpha)$ (driven by the external, time-dependent parameter $\alpha$) and the damping coefficient $\gamma$.   

%%%%%%%%%%%%%%%%%%%%
\section{II. Ratio of  Forward and Reversed Probabilities in the Low-Noise regime}
    
We here derive the path propagator for the Kramers-Moyal expansion in the weak-noise limit. We begin with  the propagator corresponding to Eq.~\eqref{eq:ss_master} for small time jumps ($\delta t \to 0$) which is given by \cite{risk89s}
    \begin{align}
          P[X|x_{-\tau}]  = \frac{1}{2\pi}\int_{-\infty}^{\infty}dk\, e^{\Gamma(k)/\epsilon}
        \label{eq:rho_IFT_saddle},
    \end{align}
    with
    \begin{align}
        \Gamma(k) &= \int_{-\tau}^{\tau }dt\,\epsilon \left(\dot{x}_t  - \epsilon a_1(x_t)\right)ik  +  \epsilon \left(\frac{\epsilon^2}{2} a_2(x_t) (ik)^{2} + \sum_{n=3}^{\infty}\frac{ (-\epsilon)^n}{n!} a_{n}(x_t) (ik)^{n}\right),
        \label{eq:G}
    \end{align}
    where we denote by $X = \{x\}^{\tau}_{-\tau}$ a trajectory of the system that starts at $t = -\tau$ and ends at $t = +\tau$, and $\alpha_{t}$ is the  external driving parameter. To obtain the conditional probability $P[X|x_{-\tau}]$ given the initial microstate $x_{-\tau}$, we apply the saddle-point method considering the low noise limit $\epsilon\to 0$ \cite{tou09s}. It is important to notice that the dependence on $x_t$ comes from the solution $k = k(x_t)$ obtained from $\partial_k\Gamma(k)=0$. Hence, from Eq.~\eqref{eq:G}, we have
    \begin{align}
        \dot{x}_t  - \epsilon a_1 = -\sum_{n=2}^{\infty} \frac{(-\epsilon)^{n}}{(n-1)!}   a_{n} (ik)^{n-1}. 
        \label{eq:k_x}
    \end{align}
    \noindent Pawula's theorem says that higher-order terms in the Kramers-Moyal expansion may violate the non-negativity of the probability distribution \cite{risk89s}. However, treating these higher orders as small perturbations, with $\epsilon^2 a_2 \gg \epsilon^{n} a_n$ for $n > 3$ allows for approximations that preserve the calculation of the moments \cite{risk89s}. To leading order, Eq.~\eqref{eq:k_x} can be approximated as $\dot{x}_t - \epsilon a_1 \approx - \epsilon^2 a_2 ik$, resulting in
    \begin{align}
        \nonumber P[X|x_{-\tau}] &\underset{\epsilon \to 0}{\sim}  \mathcal{N} \exp\left[-\frac{1}{ 2}\int_{-\tau}^{\tau}dt\, a_2\left(\frac{\dot{x}_{t} - \epsilon a_1}{\epsilon a_2}\right)^2 +\sum_{n=3}^{\infty} \frac{1}{n!}\int_{-\tau}^{\tau}dt\   a_{n}\left(\frac{\dot{x}_{t} - \epsilon a_1}{\epsilon a_2}\right)^n\right]
        \\
        \nonumber&\hspace{2mm} =  \mathcal{N} \exp\left[-\int_{-\tau}^{\tau}dt\,S(x_t, \dot{x}_t;  \alpha_t)\right]
        \\
        &\hspace{2mm} =  \mathcal{N} \exp\left[-\int_{-\tau}^{\tau}dt\,S^{G}(x_t, \dot{x}_t;  \alpha_t)\right]\exp\left[\int_{-\tau}^{\tau}dt\,S^{NG}(x_t, \dot{x}_t;  \alpha_t)\right],
        \label{eq:path_prop}
    \end{align}
    \noindent where $\mathcal{N}$ is again a normalization factor. The generalized Onsager-Machlup functionals are here given by
    \begin{align}
        &S^{G}(x_t, \dot{x}_t;  \alpha_t) = \frac{a_2}{2}\left(\frac{\dot{x}_{t} - \epsilon a_1}{\epsilon a_2}\right)^2,
        \\
        &S^{NG}(x_t, \dot{x}_t;  \alpha_t) = \sum_{n=3}^{\infty} \frac{a_{n}}{n!}  \left(\frac{\dot{x}_{t} - \epsilon a_1}{\epsilon a_2}\right)^n,
    \end{align}
    \noindent and $S = S^{G} + S^{NG}$. In order to proceed, we introduce a forward process denoted as $\alpha_t^F$, where the driving parameter is varied from an initial value $\alpha_{-\tau}^F=\alpha_i$ to a final value $\alpha^F_{+\tau} = \alpha_f$. We additionally consider the time-reversed process, denoted as $\alpha_t^R = -\alpha_{-t}^F$. Based on the expression \eqref{eq:path_prop}, the conditional probability of observing a trajectory starting at $x_{-\tau}$ for the forward process reads  %  
    \begin{align}
        P^F[X|x_{-\tau}] \underset{\epsilon \to 0}{\sim} & \mathcal{N} \exp\left[-\int_{-\tau}^{\tau}dt'\,S(x_t, \dot{x}_t;  \alpha_t^F)\right].
    \end{align}
    \noindent Analogously, the propagator for the reversed process $P^R[X^{\dagger}|x^{\dagger}_{-\tau}]$, being the time-reversed trajectory $X^{\dagger} = \{x^{\dagger} \}^{\tau}_{-\tau}$ with $x^{\dagger}_t= x_{-t} $, is
    \begin{align}
        \nonumber P^R[X^{\dagger}|x^{\dagger}_{-\tau}] \underset{\epsilon \to 0}{\sim} &\mathcal{N} \exp\left[-\int_{-\tau}^{\tau}dt'\,S(x_t^{\dagger}, \dot{x}_t^{\dagger};  \alpha_t^R)\right]
        = \mathcal{N} \exp\left[-\int_{-\tau}^{\tau}dt'\,S^{\dagger}(x_t, \dot{x}_t;  \alpha_t^F)\right],
    \end{align}
    \noindent with $S^{\dagger}(x_t, \dot{x}_t ;  \alpha_t) = S(x_t, -\dot{x}_t;  \alpha_t)$. We may now proceed to derive the fluctuation theorem for the generalized entropy production. We first evaluate the ratio of path propagators \eqref{eq:path_prop} as
  \begin{align}
        \frac{P^F[X|x_{-\tau}]}{P^R[X^{\dagger}|x^{\dagger}_{\tau}]} &= \exp\left({-\sum_{n=2}^{\infty} \frac{1 }{n!}\int_{-\tau}^{\tau }dt\,  a_n\,\mathcal{F}_n(\dot{x}_{t},x_t,\alpha_t)}\right),
    \end{align}  
    where $\mathcal{F}_n(\dot{x}_{t},x_t,\alpha_t) = \left(\left(\dot{x}_{t} - \epsilon a_1\right)^n - (-1)^n\left(\dot{x}_{t} + \epsilon a_1\right)^n\right)/(\epsilon a_2)^n$. In the low-noise limit $\epsilon \to 0$, we have $\mathcal{F}_n \sim  2n(-1)^ {n-1}\dot{x}_{t} a_1^{n-1}/(\epsilon(a_2)^n)$, since
    \begin{align}
        \nonumber & {\mathcal{F}_2
        =-\frac{2}{\epsilon}\frac{2}{a_2^2}\dot{x}_{t}  a_1}, 
        \\
        \nonumber & \mathcal{F}_3 = \frac{\dot{x}_{t}}{\epsilon a_2}\left(6 \left(\frac{ \epsilon a_1}{\epsilon a_2}\right)^2 +  2 \left(\frac{\dot{x}_t}{\epsilon a_2}\right)^2\right) 
        \underset{\epsilon \to 0}{\sim} \frac{2}{\epsilon}\frac{3}{a_2^3}\dot{x}_{t}  a_1^2, 
        \\
        \nonumber & \mathcal{F}_4  = -\frac{8}{\epsilon a_2}\left(\dot{x}_{t} \left(\frac{ \epsilon a_1}{\epsilon a_2}\right)^3 +   \epsilon a_1 \left(\frac{\dot{x}_t}{\epsilon a_2}\right)^3\right) 
        \underset{\epsilon \to 0}{\sim} -\frac{2}{\epsilon}\frac{4}{a_2^4}\dot{x}_{t}  a_1^3,
        \\
        & \nonumber\hspace{10mm} \vdots
        \\
        & \mathcal{F}_n   \underset{\epsilon \to 0}{\sim} (-1)^{n-1}\frac{2n}{\epsilon a_2^{n}} \dot{x}_{t}  a_1^{n-1}, 
    \end{align}
    \noindent where $\left| \epsilon a_1/\epsilon a_2\right| >\left|1/ \epsilon a_2\right|$, which also holds in the low-noise limit.Using  again $a_1(x_t) =-V'(x_t, \alpha_t)/\gamma$, we obtain
    \begin{align}
        \frac{P^F[X|x_{-\tau}]}{P^R[X^{\dagger}|x^{\dagger}_{\tau}]} &\underset{\epsilon \to 0}{\sim}  \exp\left[-\frac{2}{\epsilon}\int_{-\tau}^{\tau}dt\, \dot{x}_{t} \frac{V'/\gamma}{a_2}  + \frac{2}{\epsilon}\sum_{n=3}^{\infty} \frac{n}{n!}\int_{-\tau}^{\tau}dt\, \dot{x}_t \frac{ a_{n}}{ a_2^{n}}\left(V'/\gamma\right)^{n-1}\right].
         \label{eq:ratio_prop}
    \end{align}
    \noindent On the other hand, the ratio between the forward and reversed steady states \eqref{eq:steady_state} gives     
    \begin{align}
        \nonumber\frac{p_{s}(x_{-\tau}, \alpha_{-\tau})}{p_{s}(x_{\tau},  \alpha_{\tau})} = \exp\left(\Delta \phi\right) &= \exp\left(\int_{-\tau}^{\tau} dt\, \left[\dot{ \alpha}_t^F \partial_{\alpha}\phi + \dot{x}_t \partial_{x} \phi \right] \right)
        \\
        &=\exp\left(  \int_{-\tau}^{\tau} dt\, \dot{ \alpha}_t^F \partial_{ a}\phi +\frac{2}{\epsilon}\int_{-\tau}^{\tau} dt\, \dot{x}_t  \frac{V'/\gamma}{ a_2}  - \frac{2}{\epsilon}\sum_{n=3}^{\infty}\frac{2^{n-1}}{ n!}\int_{-\tau}^{\tau} dt\, \dot{x}_t \frac{ a_n}{ a_2^{n}}  \left(V'/\gamma\right)^{n-1}\right).
        \label{eq:ratio_ss}
    \end{align}
    \noindent Combining Eqs.~\eqref{eq:ratio_prop} and \eqref{eq:ratio_ss}, the ratio between the unconditional probabilities further reads
    \begin{align}
        \nonumber \frac{P^{F}(X)}{P^{R}(X^{\dagger})} &= \frac{p_{s}(x_{-\tau}, \alpha_{-\tau})P^{F}[X|x_{-\tau}]}{p_{s}(x_{\tau},  \alpha_{\tau})P^{R}[X^{\dagger}|x^{\dagger}_{\tau}]} 
        \\
        & \underset{\epsilon \to 0}{\sim} \exp(\int_{-\tau}^{\tau} dt\, \dot{ a}_t^F \partial_{ a}\phi){\exp(-\frac{2}{\epsilon}\sum_{n=3}^{\infty}\frac{(2^{n-1} + n)}{ n!}\int_{-\tau}^{\tau} dt\, \dot{x}_t\frac{ a_n(x_t)}{a_2^n(x_t)}   (V'(x_t, \alpha_{t})/\gamma)^{n-1})}.
    \end{align}

   We finally obtain the general fluctuation relation 
    \begin{align}
    \label{s33}
        \frac{P^{F}(X)}{P^{R}(X^{\dagger})} 
        &= \exp(-\int_{-\tau}^{\tau} dt\, \dot{ a}_t^F \frac{\partial_{\alpha}p_{s}}{p_{s}}) \exp(\int_{-\tau}^{\tau} dt\, \dot{x}_t \partial_x \phi^{NG})
        \\
        &{=\frac{2}{\epsilon\gamma}\int_{-\tau}^{\tau} dt\,\dot{\alpha}_t^F\frac{1}{a_2(x_t)}  V'(x_t,\alpha) - \frac{2}{\epsilon}\sum_{n=3}^{\infty}\frac{\left(2/\gamma\right)^{n-1}}{n!}\int_{-\tau}^{\tau} dt\frac{ a_n(x_t)}{a_2(x_t)^n} \left(\dot{\alpha}_t^F +\dot{ x}_t \right) V'(x_t,\alpha)^{n-1}}.
    \end{align}
    since for $n\geq 3$ and {$\gamma\gg 1$,} ${e^{-(2^{n-1} +n)/(n!\gamma^{n-1})} \sim e^{-2^{n-1}/(n!\gamma^{n-1})}}$. Equation \eqref{s33} is valid for a generic Kramers-Moyal equation and  any external potential $V(x)$ (in the weak-noise regime).

\section{III.  Numerical simulation of the non-Gaussian process}
In order to  simulate non-Gaussian processes, it is convenient to map the Kramers-Moyal equation onto a Langevin equation that is easily solved numerically. We here introduce a Langevin equation whose moments are identical to those of the Kramers-Moyal equation (11) of the main text. It has the following form \cite{gor21s}
    \begin{align}
        \gamma dx= -V'(x)dt + \sqrt{\nu}d\xi^G(t)  + \eta (d\xi_{\gamma}^P(t) + d\xi_{\nu l^2}^{*P}(t)),
        \label{eq:lang_nonG1}
    \end{align}
 where the white noise term  $\xi^G(t)$ describes a Gaussian diffusion process with unit variance, while $\xi_{z}^P(t)$ and $\xi_{z'}^{*P}(t)$ respectively represent a Poisson and a modified Poisson process with jump rates $z$ and $z'$. The modified Poisson process is taken to have zero mean and variance, $\langle d\xi_{\nu l^2}^{*P}(t)\rangle = \langle(d\xi_{\nu l^2}^{*P}(t))^2\rangle = 0$, and $\langle(d\xi_{\nu l^2}^{*P}(t))^{2n}\rangle = \nu l^2 dt$, for $n\geq 2$.
The jump amplitude, denoted as $\eta$, is independent of the system's state and follows a normal distribution with zero average  and  variance $l^{-2}$. The constant diffusion rate is moreover given by $\nu = 2\gamma kT$, with the damping coefficient $\gamma$.
   
   The Kramers-Moyal coefficients   are directly related to the jump parameters \cite{gar97s,kam07s,rei16s,risk89s} (see also Refs.~\cite{anv16s,gor21s}) via
   \begin{equation}
   a_n(x) = \lim_{{dt \to 0}} \frac{1}{dt}\langle (x(t + dt) - x(t))^n|_{x(t) = x}\rangle \sim \frac{1}{dt}\langle (dx(t))^n|_{x(t) =x}\rangle.   \end{equation}
We explicitly find for the first moment
    \begin{align}
        \left<dx(t)|_{x(t) =x}\right> &= \left<\left(-\frac{V'(x)}{\gamma} dt + \frac{\sqrt{\nu}}{\gamma}d\xi^G + \frac{\eta}{\gamma}\left(d\xi_{\gamma}^P + d\xi_{\nu l^2}^{*P}(t)\right)\right)\right> 
        \\
        &= -\frac{V'(x)}{\gamma} dt + \frac{\sqrt{\nu}}{\gamma}\left<d\xi^G \right> + \frac{1}{\gamma}\left<\eta\right> \left<(d\xi_{\gamma}^P + d\xi_{\nu l^2}^{*P}(t))\right>
        \underset{dt \to 0}{\sim} -\frac{V'(x)}{\gamma} dt,
    \end{align}
while the second moment reads
    \begin{align}
        \left<dx(t)^2|_{x(t) =x}\right> &= \left<\left(-\frac{V'(x)}{\gamma} dt +\frac{\sqrt{\nu}}{\gamma}d\xi^G + \frac{\eta}{\gamma}\left(d\xi_{\gamma}^P + d\xi_{\nu l^2}^{*P}(t)\right)\right)^2\right>
        \\
        &= \frac{(V'(x))^2}{\gamma^2} dt^2 + \frac{\nu}{\gamma^2}\left<(d\xi^G)^2 \right> + \frac{1}{\gamma^2}\left<\eta^2\right> \left<(d\xi_{\gamma}^P + d\xi_{\nu l^2}^{*P}(t))^2\right>
        \\
        &  - 2\frac{V'(x)}{\gamma}\frac{\sqrt{\nu}}{\gamma}\left< d\xi^G\right> dt - 2\frac{V'(x)}{\gamma^2} \left<\eta\right>\left<(d\xi_{\gamma}^P + d\xi_{\nu l^2}^{*P}(t))\right>dt +2\frac{\sqrt{\nu}}{ \gamma^2}\left<\eta\right>\left<d\xi^G \right>\left<(d\xi_{\gamma}^P + d\xi_{\nu l^2}^{*P}(t))\right> 
        \\
        & \hspace{-2mm}\underset{dt \to 0}{\sim} \frac{\nu}{\gamma^2}\left<(d\xi^G)^2\right> + \frac{1}{\gamma^2} \left<\eta^2\right>\left<(d\xi_{\gamma}^P)^2\right> = \left(\frac{\nu}{\gamma^2} + \frac{1}{\gamma l^2} \right) dt,
    \end{align}
    where we have used $\langle\eta\rangle = 0$ and $\langle\eta^{2n}\rangle = (2n)! /(2^n l^{2n}n!)$ \cite{anv16s,gor21s}. We moreover have (i) for  Gaussian noise,  $\langle d\xi^G\rangle = 0$, $\langle(d\xi^G)^2\rangle = dt$, and $\langle(d\xi^G)^n\rangle = 0$, for $n\geq 3$, whereas (ii) for Poisson noise  $\langle (d\xi_{z}^P)^n\rangle = z dt$ as $dt \to 0$ \cite{anv16s,gor21s}. We further  assume the independence of Wiener, Poisson, and $\eta$  processes, that is, $\langle d\xi^G \eta d\xi_{\gamma}^P \rangle = \langle d\xi^G \rangle \langle \eta \rangle \langle  d\xi_{\gamma}^P \rangle$.    
    
    The third moment  is additionally given by
    \begin{align}
        \left<dx(t)^3|_{x(t) =x}\right> &= \left<\left(-\frac{V'(x)}{\gamma} dt +\frac{\sqrt{\nu}}{\gamma}d\xi^G + \frac{\eta}{\gamma}\left(d\xi_{\gamma}^P + d\xi_{\nu l^2}^{*P}\right)\right)^3\right>
        \\
        &= -\frac{(V'(x))^3}{\gamma^3} dt^3 + \frac{\nu^{3/2}}{\gamma^3}\left<(d\xi^G)^3 \right> + \frac{1}{\gamma^3}\left<\eta^3\right> \left<(d\xi_{\gamma}^P + d\xi_{\nu l^2}^{*P})^3\right>
        \\
        &\hspace{4.1mm}  +3\left(\frac{(V'(x))^2}{\gamma^2}\frac{\sqrt{\nu}}{\gamma}\left< d\xi^G\right> dt^2 - \frac{V'(x)}{\gamma}\frac{\nu}{\gamma^2}\left< (d\xi^G)^2\right>dt\right) 
        \\
        &\hspace{4.1mm}  +3\left(\frac{(V'(x))^2}{\gamma^2}\frac{1}{\gamma}\left<\eta\right>\left<(d\xi_{\gamma}^P + d\xi_{\nu l^2}^{*P})\right> dt^2 - \frac{V'(x)}{\gamma}\frac{1}{\gamma^2}\left<\eta^2\right>\left<(d\xi_{\gamma}^P + d\xi_{\nu l^2}^{*P})^2\right> dt\right)
        \\
        &\hspace{4.1mm}  +3\left(\frac{\nu}{\gamma^2}\left< (d\xi^G)^2\right>\frac{1}{\gamma}\left<\eta\right>\left<(d\xi_{\gamma}^P + d\xi_{\nu l^2}^{*P})\right>  + \frac{\sqrt{\nu}}{\gamma}\left<d\xi^G\right>\frac{1}{\gamma^2}\left<\eta^2\right>\left<(d\xi_{\gamma}^P + d\xi_{\nu l^2}^{*P})^2\right> \right)
        \\
        &\hspace{4.1mm} - 6\frac{V'(x)}{\gamma} \frac{\sqrt{\nu}}{\gamma}\left<d\xi^G\right>\frac{1}{\gamma}\left<\eta\right>\left<\left(d\xi_{\gamma}^P + d\xi_{\nu l^2}^{*P}\right)\right>dt \underset{dt \to 0}{\sim}  0.
    \end{align}
    Finally, for $n \geq 4$, we find
    \begin{align}
        &\left<dx(t)^{2n-1}|_{x(t) =x}\right> = 0
        \\
        &\left<dx(t)^{2n}|_{x(t) =x}\right> =  \left<\eta^{2n}\right> \left<(d\xi_{\nu l^2}^P)^{2n}\right>  + \frac{1}{\gamma^{2n}} \left<\eta^{2n}\right>\left<(d\xi_{\lambda}^P)^{2n}\right> = \frac{(2n)!}{2^n n!}\left(\frac{\nu}{\gamma^2}\frac{1}{(\gamma l)^{2n-2}} + \frac{\gamma}{(\gamma l)^{2n}}\right) dt.
    \end{align}
  We may hence conclude that  all the odd Kramers-Moyal coefficients, $a_{2n-1}$ with $n \geq 2$, vanish. We therefore have
    \begin{align}
        &a_{1}(x) = -\frac{V'(x)}{\gamma}, \qquad {a_{2n}= \frac{(2n)!}{2^{n} n!}\frac{\gamma}{(\gamma l)^{2n}}\left( 1 + 2kTl^2\right)}.
    \end{align}
    which are those of the Kramers-Moyal equation (11) of the main text.


\begin{thebibliography}{99}
\bibitem{com} The proportionality factor is exactly two in the Gaussian case \cite{zon03}.
\end{thebibliography}
\end{document}